# Peritumoral Expansion Radiomics for Improved Lung Cancer Classification

**Fakrul Islam Tushar**

Dept. of Electrical & Computer Engineering, Pratt School of Engineering, Duke University, Durham.
Center for Virtual Imaging Trials, Carl E. Ravin Advanced Imaging Laboratories, Department of Radiology, Duke University School of Medicine, Durham, NC.

**Contact:** tushar.ece@duke.edu

## Abstract

**Purpose**: This study investigated how nodule segmentation and surrounding peritumoral regions influence radiomics-based lung cancer classification.

**Methods**: Using 3D CT scans with bounding box annotated nodules, we generated 3D segmentations using four techniques: Otsu, Fuzzy C-Means (FCM), Gaussian Mixture Model (GMM), and K-Nearest Neighbors (KNN). Radiomics features were extracted using the PyRadiomics library, and multiple machine-learning-based classifiers, including Random Forest, Logistic Regression, and KNN, were employed to classify nodules as cancerous or non-cancerous. The best-performing segmentation and model were further analyzed by expanding the initial nodule segmentation into the peritumoral region (2, 4, 6, 8, 10, and 12 mm) to understand the influence of the surrounding area on classification. Additionally, we compared our results to deep learning-based feature extractors Foundation Model for Cancer Biomarkers (FMCB) and other state-of-the-art baseline models.

**Results:** Incorporating peritumoral regions significantly enhanced performance, with the best result obtained at 8 mm expansion (AUC = 0.78). Compared to image-based deep learning models, such as FMCB (AUC = 0.71) and ResNet50-SWS++ (AUC = 0.71), our radiomics-based approach demonstrated superior classification accuracy.

**Conclusion**: The study highlights the importance of peritumoral expansion in improving lung cancer classification using radiomics. These findings can inform the development of more robust AI-driven diagnostic tools.

## Introduction

Lung cancer is the leading cause of cancer-related deaths worldwide, and early detection is critical for improving patient outcomes [1-3]. With the advent of artificial intelligence and radiomics, there is growing interest in leveraging quantitative imaging features to classify lung nodules as benign or malignant [2, 4-7]. However, the role of the peritumoral region in influencing classification performance remains underexplored. This study addresses these gaps by evaluating various segmentation techniques and analyzing the impact of peritumoral expansion on radionics-based classification performance.

## Methods

### Dataset

The Duke Lung Cancer Screening Dataset (DLCSD) is an open-access resource for lung cancer detection and classification research [4]. It contains over 2,000 low-dose CT scans from 1,613 patients, with more than 3,000 annotated nodules labeled with 3D bounding boxes and diagnostic outcomes (cancer or no cancer) based on the Lung-RADS lexicon. Annotations were initially performed by a virtual AI reader, and then reviewed and verified by medical students and cardiothoracic imaging radiologists [2, 4].

As shown in Table 1, the dataset features a diverse patient cohort with a mean age of 66 years and includes 50.28% male and 49.72% female participants. Patients classified as benign constitute 91.07%, while 8.93% were diagnosed as malignant. The dataset was split into training (70%), validation (20%), and testing (10%) sets. With its extensive annotations and diverse representation, DLCSD provides a robust benchmark for developing and validating AI models in lung cancer screening and classification tasks [2].

### Segmentation Techniques

To evaluate the impact of segmentation methods on radiomics feature extraction and classification, we employed four distinct techniques. Otsu thresholding is a global thresholding method that calculates an optimal threshold to minimize intra-class intensity variance, ideal for segmenting structures with distinct intensity levels [8]. Fuzzy C-Means (FCM) clustering assigns voxels to clusters based on degrees of membership, making it suitable for segmenting regions with soft boundaries or overlapping intensities [8]. The Gaussian Mixture Model (GMM), a probabilistic approach, models the dataset as a mixture of Gaussian distributions and uses the expectation-maximization algorithm to estimate voxel cluster probabilities, enabling the segmentation of complex intensity distributions [8]. Lastly, K-Nearest Neighbors (KNN) is a non-parametric method that clusters voxels based on similarity to the k-nearest neighbors in feature space, effectively segmenting heterogeneous nodule regions [8]. These segmentation masks were used for radiomics feature extraction, providing a comprehensive evaluation of threshold-based, probabilistic, and clustering-based strategies.

### Radiomics Feature Extraction

Following segmentation, radiomics features were extracted using the PyRadiomics library [9]. This included metrics capturing intensity, texture, and shape-based characteristics of the segmented nodules. These features are designed to quantitatively describe nodule properties and their surrounding structures, serving as inputs for the classification models [9].

### Classification Models

For the classification task, we employed three machine learning classifiers to evaluate the radiomics features extracted from segmented nodules. Random Forest, an ensemble-based method, was used for its ability to handle high-dimensional data and capture complex feature interactions [10]. Logistic Regression, a linear classifier, was chosen for its interpretability and robust performance in binary classification tasks [11]. Lastly, KNN, a non-parametric method, classified nodules based on their proximity to neighboring samples in the feature space. Each classifier was trained on radiomics features derived from different segmentation methods, and their performance was assessed. This multi-classifier approach enabled a comprehensive evaluation of the effectiveness of radiomics features in distinguishing between cancerous and non-cancerous nodules.

Additionally, we compared these models to previously developed deep learning-based classifiers from earlier studies [4, 6]. These included five baseline models, ranging from randomly initialized ResNet50 to pretrained and self-supervised approaches such as the **Foundation Model for Cancer Biomarkers (FMCB)**, Models-Genesis, MedNet3D, and our in-house **ResNet50-SWS**++ [4, 6]. Among these, the SWS++ model was specifically designed with a Strategic Warm-Start pretraining approach to enhance lung nodule classification performance. All these models processed 64 × 64 × 64 (x, y, z) patches of nodules with identical training configurations. This inclusion allowed us to comprehensively compare the performance of radiomics-based models with deep learning-based approaches, emphasizing the impact of contextual tissue information in lung cancer classification.

### Peritumoral Expansion

To assess the influence of the surrounding lung tissue, the initial segmentation masks were systematically expanded into the peritumoral regions by distances of 2, 4, 6, 8, 10, and 12 mm. Radiomics features were then re-extracted from these expanded regions to analyze the impact on cancer/no-cancer classification performance.

### Evaluation Metrics

The performance of each segmentation and classification combination was evaluated using the Area Under the Curve (AUC) of the Receiver Operating Characteristic (ROC) curve [12]. This metric was used to quantify the discriminative ability of the models, with higher AUC values indicating better classification performance.

## Results & Discussion

**Fig. 1** illustrates examples of the generated segmentation using KNN and the peritumoral expansion. **Table 2** presents the AUC-ROC results for Random Forest, Logistic Regression, and KNN models across the four segmentation techniques over the validation dataset. Logistic Regression paired with the KNN segmentation method achieved the best performance, with an AUC of 0.89. Random Forest also performed well with the KNN segmentation, achieving an AUC of 0.88, slightly lower than Logistic Regression. The Otsu and GMM segmentation methods produced comparable AUC values, with Logistic Regression consistently outperforming other classifiers. FCM segmentation yielded the lowest classification performance across all classifiers, with AUC values ranging from 0.75 to 0.84. The best-performing segmentation (KNN) and model (Logistic regression) were further analyzed by expanding the initial nodule segmentation into the peritumoral region (2, 4, 6, 8, 10, and 12 mm) to understand the influence of the surrounding area on classification.

### Analysis of Peritumoral Expansion on Classification Performance

In our study, we extended the segmentation regions to include peritumoral tissue and assessed the impact on lung cancer classification performance. As depicted in the fig. 2(b), expanding the segmentation region (**initial KNN segmentation**) around the nodule resulted in progressive improvements in classification AUC. Starting from the original nodule segmentation (**AUC = 0.73; 95% CI: 0.63–0.82**), incorporating a **2 mm peritumoral region** maintained a similar AUC (**0.73;0.62–0.83**). Expanding the region further to **4 mm** improved the AUC to **0.75 (0.64–0.85)**, while **6 mm** and **8 mm expansions** yielded the highest performance with AUCs of **0.77 (0.68–0.86)** and **0.78 (0.69–0.87)**, respectively. However, beyond the 8 mm expansion, the performance plateaued, with a **10 mm expansion** resulting in an AUC of **0.76 (0.66–0.84)** and a slight decline observed for the **12 mm expansion** (**0.70;0.58–0.81**). These results indicate that incorporating peritumoral regions up to 8 mm provides critical diagnostic information that enhances the model's ability to classify nodules, likely capturing relevant tissue characteristics associated with malignancy. Beyond this threshold, the inclusion of additional peritumoral tissue appears to introduce noise, negatively impacting model performance.

### Comparison with Deep Learning Models

When comparing our peritumoral analysis with image-based deep learning models (**fig. 4**). The best-performing deep learning model **ResNet50-SWS**++, achieved an AUC of **0.71 (0.61–0.81)**, which is lower than our radiomics-based Logistic Regression model with peritumoral expansion up to 8 mm (**0.78;0.69–0.87) [4]**. In lung cancer classification, FMCB served as a feature extractor, with its features used to train a Logistic Regression model for distinguishing cancerous from non-cancerous nodules [4]. This hybrid approach combines the features of deep learning with the simplicity of Logistic Regression. FMCB achieved an AUC of 0.71(0.60–0.82), but radiomics-based methods with peritumoral expansion outperformed it, highlighting the importance of contextual tissue information in diagnosis. Models such as Genesis (**AUC = 0.64**) and MedNet3D (**AUC = 0.67**) also underperformed compared to our approach with optimized peritumoral inclusion [4]. This comparison highlights the advantage of leveraging radiomics features and systematically expanding the segmentation to incorporate contextual tissue information. While deep learning models excel in capturing image-based features, they may not fully utilize the microenvironment surrounding nodules, as radiomics-based approaches do. These findings suggest that combining peritumoral analysis with machine learning could complement deep learning techniques, offering a comprehensive framework for improving lung cancer diagnosis.

This study demonstrates the critical role of the inclusion of peritumoral regions in improving radiomics-based lung cancer classification. The findings suggest that incorporating surrounding tissue up to 8 mm provides valuable diagnostic information, likely capturing tumor microenvironment characteristics. These insights can be used to refine AI models for lung cancer screening and diagnosis. Future work could explore integrating these approaches into end-to-end deep learning pipelines or applying them to larger, more diverse datasets.

## Conclusion

Our study underscores the importance of segmentation techniques and peritumoral analysis in lung cancer classification. By leveraging optimized segmentation and radiomics features, we provide a framework for developing robust AI-based diagnostic tools that can improve early detection and patient outcomes.

## Acknowledgment

This work was funded in part by the Center for Virtual Imaging Trials, NIH/NIBIB P41-EB028744 and Putnam Vision Award awarded by Duke Radiology. I also suggest acknowledging the Duke Lung Cancer Screening Program.

## Dataset and Code Availability

The dataset used in this study is the publicly available Duke Lung Cancer Screening Dataset (DLCSD) at Zenodo: https://zenodo.org/records/10782891. The code for data preprocessing, segmentation, feature extraction, model training, and evaluation will be openly available at GitHub: https://github.com/fitushar/AI-in-Lung-Health-Benchmarking-Detection-and-Diagnostic-Models-Across-Multiple-CT-Scan-Datasets

**Table 1. Demographic distribution of the data Cohort used for training, development and test sets.**

| Category | | All (%) | Training (%) | Validation (%) | Testing (%) |
|---|---|---|---|---|---|
| Gender | | | | | |
| | Male | 811 (50.28) | 559 (52.48) | 167 (46.78) | 85 (42.93) |
| | Female | 802 (49.72) | 499 (47.16) | 190 (53.22) | 113 (57.07) |
| Age | Mean (min-max) | 66 (50-89) | 66 (50-89) | 66 (55-78) | 66 (54-79) |
| Race | White | 1,195 (74.09) | 775 (73.25) | 280 (78.43) | 140 (70.71) |
| | Black/AA | 366 (22.69) | 247 (23.35) | 68 (19.05) | 51 (25.76) |
| | Other/Unknown | 52 (3.22) | 36 (3.40) | 9 (2.52) | 7 (3.54) |
| Ethnicity | | | | | |
| | Not Hispanic | 1,555 (96.40) | 1,019 (96.31) | 344 (96.36) | 192 (96.97) |
| | Unavailable | 52 (3.22) | 35 (3.31) | 12 (3.36) | 5 (2.53) |
| | Hispanic | 6 (0.37) | 4 (0.38) | 1 (0.28) | 1 (0.51) |
| Smoking status | | | | | |
| | Current | 826 (53.92) | 538 (53.48) | 189 (56.08) | 99 (52.38) |
| | Former | 704 (45.95) | 467 (46.42) | 147 (43.62) | 90 (47.62) |
| | Other/Unknown | 2 (0.13) | 1 (0.10) | 1 (0.30) | |
| Cancer | | | | | |
| | | Patient | | | |
| | Benign | 1,469 (91.07) | 965 (91.21) | 324 (90.76) | 180 (90.91) |
| | Malignant | 144 (8.93%) | 93 (8.79) | 33 (9.24) | 18 (9.09) |
| | | Nodule | | | |
| | Benign | 2,223 (89.38) | 1,452 (89.74) | 510 (88.70) | 261 (88.78) |
| | Malignant | 264 (10.62) | 166 (10.26) | 65 (11.30) | 33 (11.22) |

**Table 2: AUC Performance for Different Segmentation and Classification Combinations over the Validation Dataset.**

| Segmentation Method | Random Forest | Logistic Regression | KNN |
|---|---|---|---|
| Otsu | 0.85 | 0.86 | 0.78 |
| FCM | 0.82 | 0.84 | 0.75 |
| KNN | 0.88 | 0.89 | 0.81 |
| GMM | 0.83 | 0.85 | 0.77 |

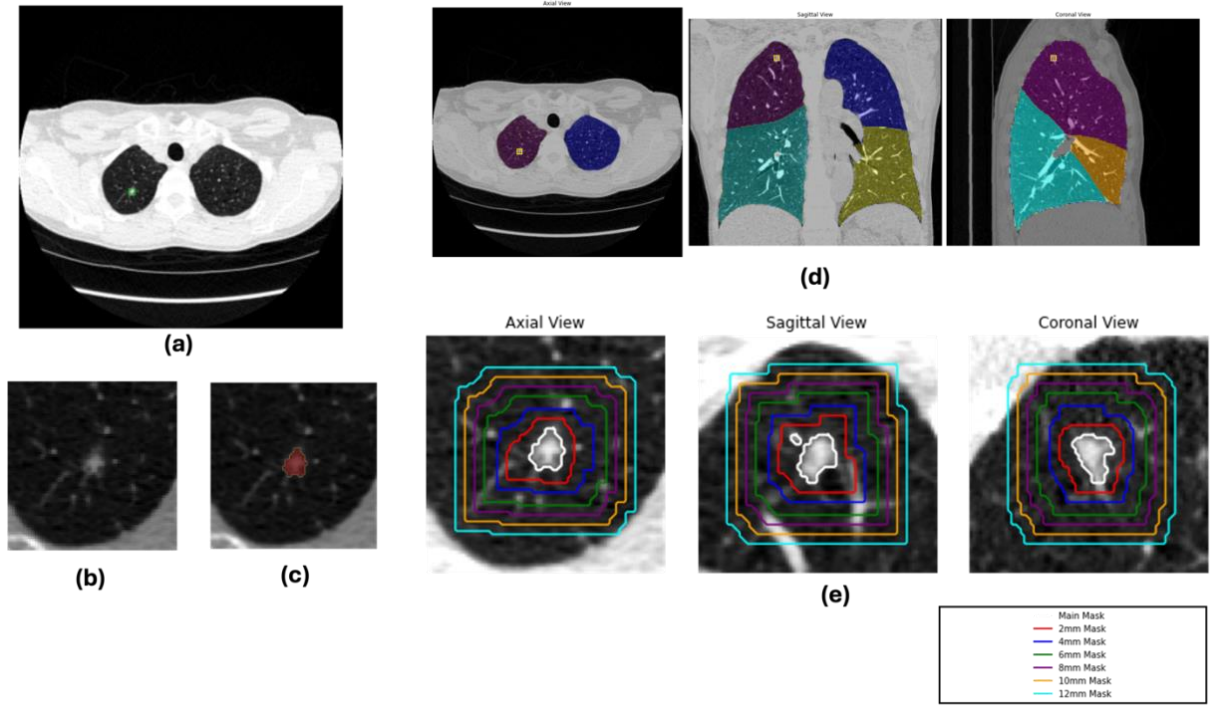

**Figure 1. Illustrates the segmentation process and visualization of a lung nodule using the K-Nearest Neighbors (KNN) segmentation technique.** **(a)** Displays an axial view of the CT scan showing the nodule's location, highlighted with a green bounding box. **(b)** A magnified view of the nodule region, and **(c)** demonstrates the segmented nodule overlaid in red, highlighting the precise delineation achieved by KNN segmentation. **(d)** Multi-planar reconstructions of the segmented lung, showcasing axial, sagittal, and coronal views with segmented lung regions color-coded to represent anatomical divisions. This segmentation helps in localizing nodules with respect to lung regions for further analysis. **(e)** depicts the peritumoral segmentation strategy, where concentric masks around the nodule expand by increments of 2 mm (up to 12 mm). Each mask is represented by a different color. These expanded regions capture the peritumoral tissue, which is critical for assessing the impact of surrounding tissue on classification performance.

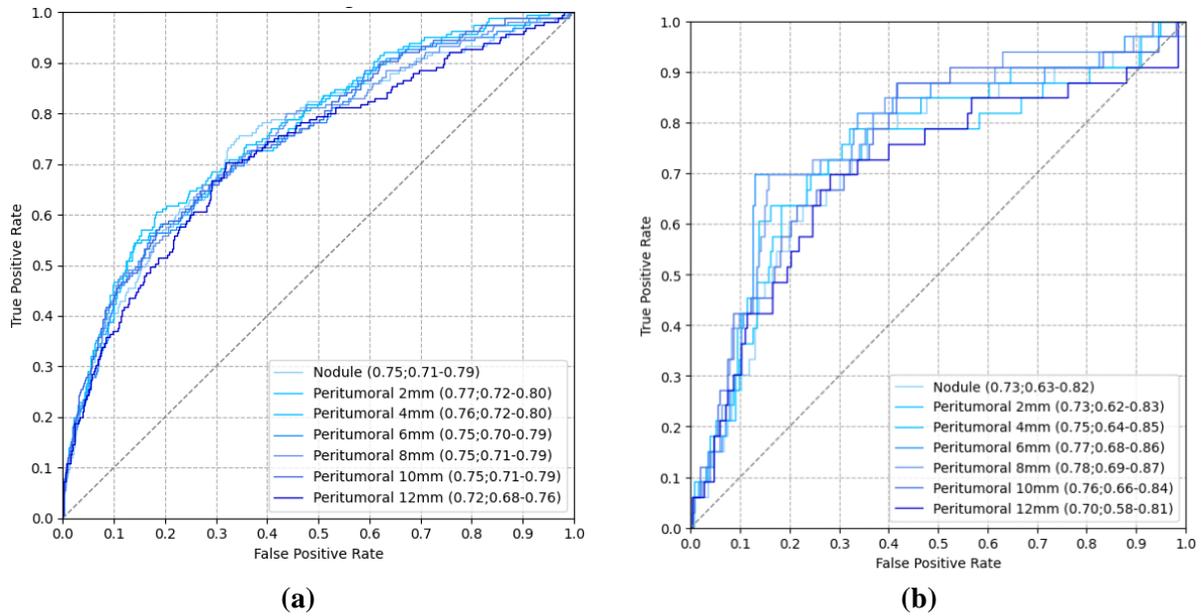

**Figure 3. (a) Training and (b) testing results of Logistic Regression with expanded segmentation.**

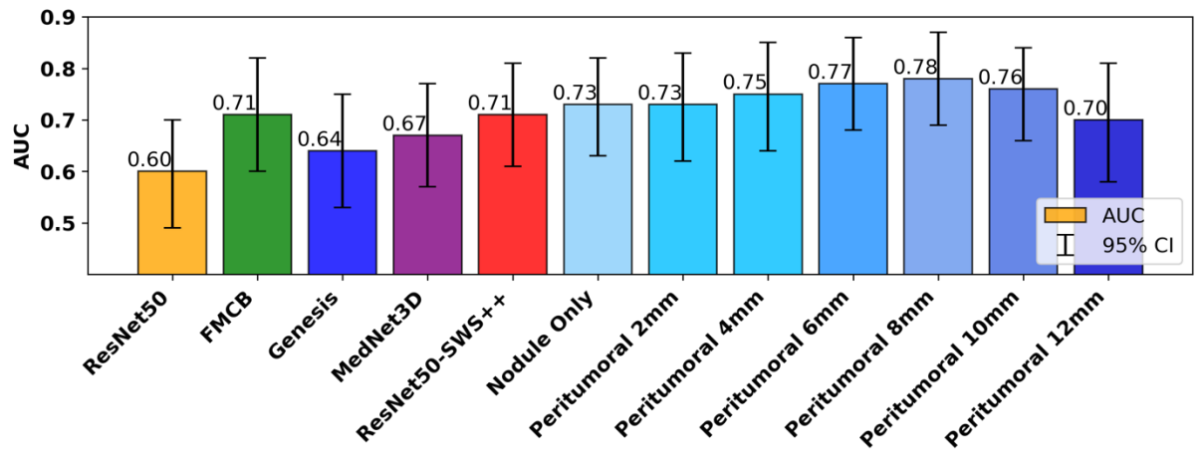

**Figure 4. AUC Comparison of Deep Learning and Radiomics-Based Models with 95% Confidence Intervals.** The deep learning models include ResNet50, FMCB (feature extractor with Logistic Regression), Genesis, MedNet3D, and ResNet50-SWS++ [4]. The radiomics-based models utilize Logistic Regression trained on features extracted from nodules and expanded peritumoral regions (2 mm to 12 mm). Bars represent the mean AUC values, with error bars indicating the 95% confidence intervals for each model. Radiomics-based models with peritumoral expansion (e.g., 6 mm and 8 mm) demonstrate higher classification performance compared to deep learning models. The best radiomics-based result, achieved with an 8 mm expansion, outperformed all deep learning approaches, highlighting the value of incorporating peritumoral regions in lung cancer classification.

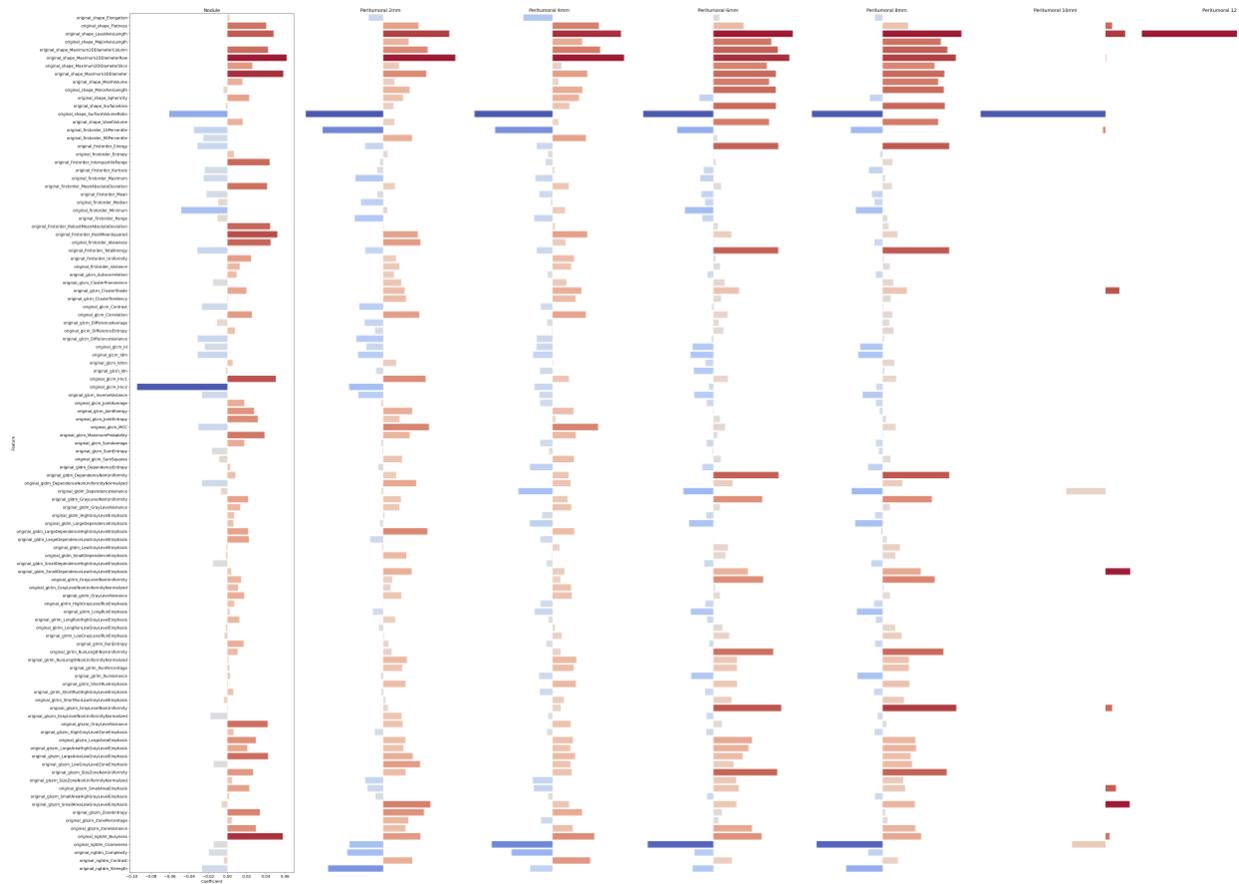

**Figure 5.** Radiomics feature importance.